\documentclass[preprint2]{aastex}

\usepackage{graphicx,rotate}
\usepackage{natbib}
\usepackage{latexsym,amssymb,verbatim}

\shorttitle{Molecular tracers of PDR}
\shortauthors{Bayet et al.}

\begin{document}


\title{Molecular tracers of PDR-dominated galaxies}


\author{E. Bayet\altaffilmark{1}, S. Viti\altaffilmark{1},
D.A. Williams\altaffilmark{1}, J.M.C. Rawlings\altaffilmark{1}
and T. Bell\altaffilmark{2}}

\email{eb@star.ucl.ac.uk}


\altaffiltext{1}{Department of Physics and Astronomy, University
College London, Gower Street, London WC1E 6BT, UK.}
\altaffiltext{2}{Department of Astronomy, California Institute of
Technology, Pasadena, CA 91125, USA}


\begin{abstract}
Photon-dominated regions (PDRs) are powerful molecular line
emitters in external galaxies. They are expected in galaxies with
high rates of massive star formation due to either starburst (SB)
events or starburst coupled with active galactic nuclei (AGN)
events. We have explored the PDR chemistry for a range of physical
conditions representing a variety of galaxy types. Our main result
is a demonstration of the sensitivity of the chemistry to changes
in the physical conditions. We adopt crude estimates of relevant
physical parameters for several galaxy types and use our models to
predict suitable molecular tracers of those conditions. The set of
recommended molecular tracers differs from that which we
recommended for use in galaxies with embedded massive stars. Thus,
molecular observations can in principle be used to distinguish
between excitation by starburst and by SB+AGN in distant galaxies.
Our recommendations are intended to be useful in preparing
Herschel and ALMA proposals to identify sources of excitation in
galaxies.
\end{abstract}






\section{Introduction}\label{sec:intro}

Galaxies may be powerful molecular line emitters because of a very
high massive star formation rate induced either by starburst (SB)
events or the presence of active galactic nuclei
(AGN)\citep{Radf91, Solo92, Geri01, Isra01,Isra02, Isra03, Baye04,
Solo04, Mart05, Baye06, Evan06, Grac06, Mart06, Aalt07c, Bell07,
Iman07, Knud07, Papa07, Krip08}. Some external sources may be
dominated by only one type of activity or contain strong
signatures of both. Both activity types (SB and AGN) are
recognized as having a direct influence on the molecular gas
emission in a galaxy.

Molecular emission from dense star-forming cores in starburst
galaxies has recently been the subject of a modelling study
\citep{Baye08a}. That work predicts that such galaxies should be
strong emitters in rotational lines from molecules embedded in the
warm cores from which the young massive stars are evolving (see
also \citealt{Lint05, Lint06}). \citet{Baye08a} described how the
chemistry in these cores may depend on the local physical
conditions, and they identified some molecular species that should
be sensitive tracers of these conditions.

Such starburst galaxies may also contain giant photon-dominated
regions (PDRs) which are also powerful emitters in molecular lines
in the millimetre and submillimetre regions of the spectrum. These
PDRs are quite distinct physical and chemical systems from those
discussed by \citet{Baye08a}. Giant PDRs should be set up in
galaxies with clusters of young massive stars and in galaxies
dominated by radiation from an active nucleus. It is therefore
important to discuss the chemistry of galaxies dominated by PDRs,
set up either by clusters of young massive stars or by the
additional presence of active galactic nuclei (AGN), and to
identify the likely molecular tracers of such PDRs. In particular,
it would be important to know whether these molecular signatures
are similar to or distinct from those of the star-forming cores of
starburst galaxies described by \citet{Baye08a}.

In this companion paper to \citet{Baye08a}, we model the chemistry
of PDR-dominated galaxies. As described in the earlier paper, the
physical conditions that determine the chemistry in the
interstellar medium of external galaxies (metallicity and/or
relative elemental abundances; dust:gas ratio; dust grain
properties; gas density; cosmic ray ionisation rate; and far UV
radiation intensity) may differ substantially from those of the
Milky Way and for our purposes are regarded as unknown parameters.
The chemistry is explored for a wide range of these parameters and
its sensitivity to variations in these parameters is studied. We
also attempt to assign likely values of these parameters to galaxy
type (starburst, SB+AGN and high-redshift) and predict suitable
molecular tracers for these galaxy types.

The PDR code and the chemical database are described in Section
\ref{sec:mod}, and the physical parameters and their ranges are
discussed in Section  \ref{sec:para}. Results describing the PDR
chemistry and its sensitivity to variations in the parameters
listed above are presented in Section \ref{sec:sensi}. In Section
\ref{sec:gal} we make predictions of the detectability of lines of
certain potential tracer molecules in PDR-dominated galaxies at
low and high redshift, and in Section \ref{sec:conclu} we give a
discussion of our results in relation to observations, to other
PDR modelling, and to our earlier warm-core studies, and we
summarize our conclusions.

\section{Model Description}\label{sec:mod}

PDR models differ from each other in the assumptions adopted, in
the geometry, or in the degree of sophistication used for solving
the radiative transfer equations, or in the thermal balance and
for determining the level population. PDR models also differ from
each other by the chemical network adopted and by the
consideration (or not) of time-dependence. In this paper, we used
the time-dependent UCL\_PDR code benchmarked in \citet{Roel07} and
fully described in \citet{Bell06, Bell07}.

The UCL\_PDR model assumes that a one-dimensional semi-infinite
slab is illuminated from one side by FUV photons. The chemistry
and thermal balance are calculated self-consistently at each point
into the slab and at each time-step, producing chemical
abundances, emission line strengths and gas temperatures as a
function of depth and time. The UCL\_PDR code adopted here uses a
chemical network containing 131 species and over 1700 reactions
from the UMIST99 database \citep{LeTe00}, including ion-molecule,
photoionisation and photodissociation reactions. The UCL\_PDR
model also includes some modifications of UMIST99 introduced as
part of the PDR benchmarking effort presented in \citet{Roel07}.
The freeze-out of atoms and molecules on to grains is neglected in
this model. Apart from hydrogen (which is assumed to be initially
mainly molecular), the gas is assumed to be initially in a purely
atomic form, with initial elemental abundance ratios as free
parameters. Initial elemental abundances of all metals are assumed
in most cases to scale linearly with the metallicity
(z/z$_{\odot}$), and the H$_{2}$ formation rate and the adopted
gas-to-dust mass ratio are also assumed to be linear with the
metallicity. For a more detailed description of the UCL\_PDR
model, we refer the reader to \citet{Bell06, Bell07} and
references therein.

\section{Parameter choices}\label{sec:para}

To perform our study, we used and extended a model grid previously
built and presented by \citet{Bell06, Bell07}. This grid contains
more than 1200 models, each with a different set of input
parameters. We restricted our selection to 9 models, the most
relevant for our study. Table \ref{tab:stand} lists the standard
(Milky Way) values of the parameters. Table \ref{tab:model_grid}
lists the range of parameter values covered by the models we have
selected. We adopt solar initial chemical abundances as our
standard. Models 9, 10 and 11 explore the effect of replacing the
solar initial elemental abundance ratios by values predicted by
early universe models from \citet{Chie02, Hege02, Umed02} (see
Table \ref{tab:ele_ab_ratios}). In these three models, the
metallicity, the H$_{2}$ formation rate and the gas-to-dust mass
ratio are fixed to standard values. As described later in Sect.
\ref{sec:gal}, Model 12 is intended to represent high-redshift
environments.

Models 0 to 4 investigate the consequences of a reduction of solar
metallicity by a factor up to 100 consistently linked with a
reduction of the H$_{2}$ formation rate, of the gas-to-dust ratio
and of the initial elemental abundance ratios. Models 0 and 6
study the consequent changes in the chemistry after an increase of
cosmic ray ionization rate from $5 \times 10^{-17}$s$^{-1}$ to $5
\times 10^{-15}$s$^{-1}$. We suggest that the increase of cosmic
ray ionization rate may be qualitatively used to simulate XDR-like
environments. Finally, a two orders of magnitude change in both
the FUV radiation field (Models 0 and 7) and in the gas density
(Models 0 and 8) have been investigated.

Each model provides atomic and molecular abundances as a function
of total cloud depth (at visual extinctions of 0 $<$ A$_{v}<$ 10)
for cloud ages of 10$^{4}$, 10$^{5}$, 10$^{6}$, 10$^{7}$ and
10$^{8}$yr. For the study of the A$_{v}$ influence (Models 0 and 5
in Table \ref{tab:model_grid}), we have restricted our model
selection to those showing either a value of A$_{v}$=8, or a value
of A$_{v}$ = 3 (typical of a translucent gas component). Note that
Model 0 is considered as our reference model in the paper,
assuming standard parameters (listed in Table \ref{tab:stand})
with an A$_{v} = $ 8, likely to be a typical value for
representing the dense PDR gas component in galaxies, as detected
in the nuclei of M~82, NGC~253, IC~342, and NGC~4038
\citep{Baye08b, Baye08c}.

\begin{table*}
    \caption{Standard model parameters \citep{Bell06, Bell07} and
    references therein. These parameters are those used in
    Models 0-12 except if a different value is specified
    in Table \ref{tab:model_grid} or described in the
    text.}\label{tab:stand}
    \begin{tabular}{l r}
    \hline
    Grain size & 0.1 $\micron$\\
    Grain albedo & 0.7 \\
    Microturbulence velocity & 1.5 kms$^{-1}$\\
    Mean photon scattering by grain & 0.9\\
    External FUV radiation intensity (=I$_{\odot}$) & 1 Habing$^{a}$ \\
    Cosmic ray ionization rate (=$\zeta_{\odot}$) & 5.0$\times 10^{-17}$s$^{-1}$\\
    H$_{2}$ formation rate coefficient (=R$_{\odot}$) & 3$\times 10^{-18}\sqrt{T} \exp(\frac{-T}{1000})$ cm$^{3}$s$^{-1}$\\
    Gas:dust mass ratio (=d$_{\odot}$) & 100 \\
    Metallicity (=z$_{\odot}$) & solar values $^{b}$\\
    Si/H & 8.21 $\times 10^{-7}$\\
    Fe/H & 3.60 $\times 10^{-7}$\\
    Cl/H & 1.10 $\times 10^{-7}$\\
    Na/H & 8.84 $\times 10^{-7}$\\
    Ca/H & 5.72 $\times 10^{-10}$\\
    \hline
    \end{tabular}

    $^{a}$: The unit of the standard  Interstellar Radiation Field (ISRF) intensity
    is I$_{\odot}$ = 1.6$\times 10^{-3}$erg cm$^{-2}$s$^{-1}$ \citet{Habi68};
    $^{b}$ : z = 1 = z$_{\odot}$ corresponds to solar values of the initial
    elemental abundance ratios (see Table \ref{tab:ele_ab_ratios}) while
    z = 1/10 z$_{\odot}$ means that the solar values of the initial elemental
    abundance ratios have been all divided by the same factor (of 10 in this
    example).
\end{table*}

\section{Sensitivity of chemical abundances to variations of the
UCL\_PDR model parameters}\label{sec:sensi}

In this section, we present results on the trends in the chemistry
with respect to the changes of various parameters. We have
selected 19 molecules for closer study, either for their likely
detectability or for their chemical interest (or both). This set
of 19 molecules is closely similar to the set of molecules used in
\citet{Baye08a}. We use this set to make a comparison between the
predictions of the PDR model and those of models describing the
chemistry of high-mass star-forming regions (see Sect.
\ref{sec:conclu}). We have arbitrarily fixed the limit of
detectability of molecule X to be a relative abundance of
n(X)/n$_{H}=1\times 10^{-12}$ (where n$_{H}=$n(H) $+$
2n(H$_{2}$)), a criterion roughly satisfied in our own galaxy. The
article deals specifically with molecules CO, H$_{2}$O, CN, OH,
CS, HNC, HCN, HCO$^{+}$, H$_{3}$O$^{+}$, SO, C$_{2}$ and C$_{2}$H
(see Figs. \ref{fig:met} to \ref{fig:ele_ab_ratios}) and with
CO$_{2}$, OCS, SO$_{2}$, H$_{2}$S, H$_{2}$CS, CH$_{2}$CO and
H$_{2}$CO (see Tables \ref{tab:trend_meta} -
\ref{tab:trend_dens}). Figs. \ref{fig:met} to \ref{fig:dens} show
the time-evolution of molecular abundances and their sensitivity
to parameter changes. Unlike for lower density models
\citep{Bell06}, the chemical steady-state is achieved here in
$\approx 10^{5}$yr. Thus, time-dependence is unlikely to play a
role in the chemistry.

\subsection{Metallicity}\label{subsec:meta}

\begin{figure}
    \includegraphics[width=8.7cm]{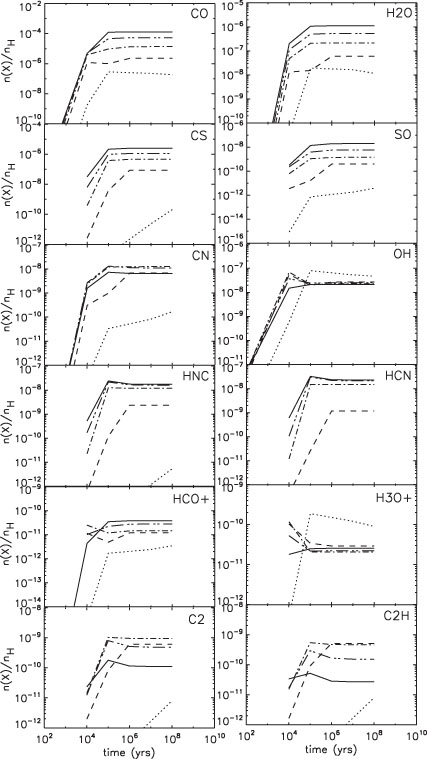}
    \caption{Influence of metallicity variations on some molecular
    abundances. We plot in a logarithmic scale the relative abundances
    n(X)/n$_{H}$ of various species, X, with respect to the time (in
    yrs). Results from the first five models in Table \ref{tab:model_grid}
    (Models 0 to 4) are plotted in this figure: Model 0 (solar metallicity)
    is represented by black lines, Model 1 (1/2 z$_{\odot}$) by
    dash-two-dots lines while Models 2 (1/4 z$_{\odot}$),
    3 (1/10 z$_{\odot}$) and 4 (1/100 z$_{\odot}$) are symbolized
    by dash-one-dot lines, dashed lines and dotted lines, respectively,
    showing the effect of decreasing the metallicity.}\label{fig:met}
\end{figure}

Fig. \ref{fig:met} shows that the model with the lowest
metallicity (Model 4 in Table \ref{tab:model_grid}) predicts very
low abundances for all species except CO, H$_{2}$O, HCO$^{+}$ and
OH, and enhances significantly the abundance of H$_{3}$O$^{+}$.
This enhancement is caused by a severe reduction in the electron
density at low metallicity. Since the main loss of polyatomic ions
is through dissociative recombination, the ion abundance
increases.

With z$/$z$_{\odot}$=0.01, we expect that the deuterated ions such
as H$_{2}$D$^{+}$ and H$_{2}$DO$^{+}$ may be more readily
detectable than H$_{3}$O$^{+}$ (see e.g. \citealt{Bore93, Paga07}
and references therein). The deuteration in both PDR and high-mass
star forming models will be investigated in a forthcoming paper.

For models having a metallicity equal to or greater than 1/10
z$_{\odot}$ (Models 0, 1, 2 and 3 in Table \ref{tab:model_grid}),
we can separate species into three different categories: (a)
species with relative abundances varying linearly with the
metallicity changes e.g. CO, H$_{2}$O, CS, SO (see Fig.
\ref{fig:met}) and CO$_{2}$, SO$_{2}$, OCS and H$_{2}$S (see Table
\ref{tab:trend_meta}); (b) molecules such as CN, OH, HCN and HNC
showing chemical abundances that are rather insensitive to the
metallicity changes; and (c) species whose abundances changes are
inversely dependent on z, such as C$_{2}$, C$_{2}$H (see Fig.
\ref{fig:met}), and H$_{2}$CO, CH$_{2}$CO and H$_{2}$CS (see Table
\ref{tab:trend_meta}).

Such varieties of behavior have been found in the earliest studies
of chemical sensitivity to parameters \citep{Pick81} and in recent
work \citep{Baye08a} and it is reasonably well understood.
Category (a) species may be lost to processes independent of
metallicity (e.g. CO is lost to dissociative ionization with
He$^{+}$)

 while category (b) species are both formed and destroyed
in exchange reactions whose reactants depend on metallicity and
are therefore largely neutral to metallicity changes (For example,
OH depends mainly on oxygen atoms for its formation and on
exchange reactions with O and other atoms for its loss). Formation
schemes for some category (c) species depend on the production of
C$^{+}$ ions from CO, a molecule proportional to z, but loss by a
succession of exchange reactions resulting in a strong inverse
dependence on z; thus, there is an overall inverse dependence of
the abundance of such species on z.

\subsection{Initial elemental abundance
ratios}\label{subsec:ele_ab_ratios}

\begin{figure}
    \includegraphics[width=8.7cm]{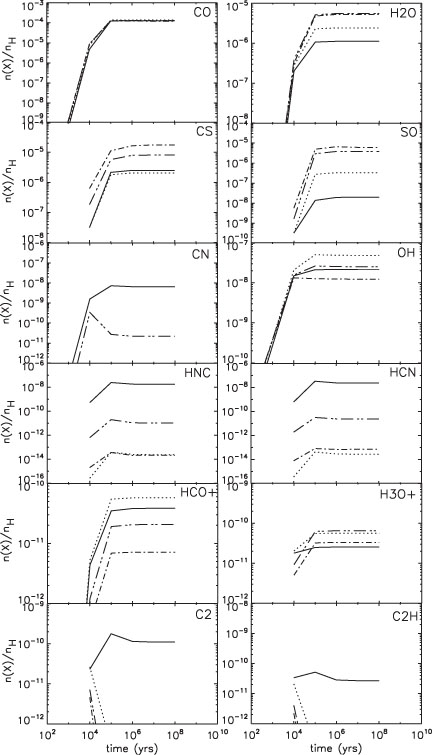}
    \caption{Influence of initial elemental abundance ratios variations
    on some molecular abundances. We plot in a logarithmic scale the
    relative abundances n(X)/n$_{H}$ of various species, X, with respect
    to the time (in yrs). Only the solar (Model 0) and the last three
    models from Table \ref{tab:model_grid} (Models 9 to 11) are plotted.
    Model 0 is represented by black lines, Model 9 by dotted lines while
    Models 10 and 11 are symbolized by dash-one-dot lines, dash-two-dots
    lines, respectively. The values of the initial elemental abundance
    ratios for each model are given in Table \ref{tab:ele_ab_ratios}.
    }\label{fig:ele_ab_ratios}
\end{figure}

We have investigated the consequences of making changes in the
initial elemental abundance ratios using Models 9, 10 and 11 (see
Table \ref{tab:model_grid}). These three models correspond to a
study of the influence of a decrease of nitrogen (Model 9), and an
increase of both the oxygen and sulfur content (Model 10 and 11).
The values of the corresponding initial elemental abundance ratios
are listed in Table \ref{tab:ele_ab_ratios}. Model 9 refers to
values from \citet{Chie02} while Model 10 and 11 used the values
suggested by \citet{Hege02} and \citet{Umed02}, respectively.
Several time-dependent chemical trends are shown in Fig.
\ref{fig:ele_ab_ratios}.

Neither oxygen- and sulfur-rich environments nor nitrogen-poor
environments strongly affect the values of the abundances of CO,
OH, H$_{2}$S, HCO$^{+}$, H$_{2}$O and H$_{3}$O$^{+}$. Indeed,
after 10$^{5}$yr, the scatter in their relative abundances is
small, within a factor of $\leq$ 10. The same behavior is seen for
the CO$_{2}$ relative abundance. These species are thus likely to
be irrelevant as tracers of initial metal gas content. On the
contrary, SO and CS (see Fig. \ref{fig:ele_ab_ratios}) as well as
OCS and SO$_{2}$ abundances (see Table
\ref{tab:trend_ele_ab_ratios}) are maximum for the oxygen- and
sulfur-rich models and the variation of their relative abundances
is important (within a factor of $<$ 100). This makes them very
sensitive tracers. For Model 0, it is shown that CN, HNC, HCN,
C$_{2}$ and C$_{2}$H reach their highest relative abundances
having in addition significant variations (factor $\gtrsim
10^{3}$) as compared to Model 9 (nitrogen-poor model). These last
two species are seen particularly sensitive to the nitrogen gas
content since they are likely to only survive in nitrogen-rich
environment (for a N/H ratio higher than 6.5$\times 10^{-5}$). The
models having the highest O/H and S/H ratios do not reveal
H$_{2}$CO, H$_{2}$CS and CH$_{2}$CO as particularly good tracers.
Their corresponding relative abundances reach their lowest values
for Models 10 and 11 while their maximum correspond to Model 0.
The decrease in relative abundance seen for some of them when the
O/H and S/H elemental abundance ratios are increased, is so
pronounced (e.g. up to a factor of 1000 for H$_{2}$CO) that the
abundances fall under the limit of detectability. These molecules
are thus likely to be undetectable in oxygen- and sulfur-rich
primordial gas while potentially observable in nitrogen-rich
environments. The trends we have identified are summarized in
Table \ref{tab:trend_ele_ab_ratios}.

\subsection{A$_{v}$}\label{subsec:Av}

\begin{figure*}
     \centering
     \includegraphics[width=12cm]{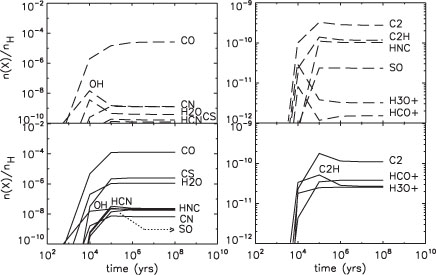}
     \caption{Influence of A$_{v}$ variations on some molecular abundances.
     We plot in a logarithmic scale the relative abundances n(X)/n$_{H}$ of
     various species,  X, with respect to the time (in yrs). The upper two
     panels (dashed lines) are dedicated to Model 5 (A$_{v}$ = 3 in Table
     \ref{tab:model_grid}) while the lower two plots (black lines) present
     the results of the solar model (Model 0 in Table \ref{tab:model_grid})
     for A$_{v}$ = 8.}\label{fig:Av}
\end{figure*}

As expected, and shown on Fig. \ref{fig:Av}, the abundances of
many molecular species are much smaller at an A$_{v}$ = 3 (Model 5
in Table \ref{tab:model_grid}) than at A$_{v}$ = 8 (Model 0 in
Table \ref{tab:model_grid}). For example, H$_{2}$O, HNC and SO
molecules show fractional abundances smaller by more than two
orders of magnitude at A$_{v}$ = 3 than at A$_{v}$ = 8. For CS and
SO$_{2}$, the reduction factors are even larger (see Fig.
\ref{fig:Av} and Table \ref{tab:trend_Av}). These species are very
sensitive to  A$_{v}$ and good tracers of opacity. At A$_{v}=$ 3,
however, some species become undetectable since their
corresponding relative abundance drop below the probable limit of
detectability.

The strong response to A$_{v}$ is due to the ability of FUV
photons to penetrate the translucent cloud conditions (A$_{v}
\approx 3$) but not much deeper. CO is self-shielding and
generally insensitive for A$_{v} \gtrsim 1$. Some other molecules
(C$_{2}$, C$_{2}$H) depend on dissociative ionization of CO, and
so are also fairly insensitive to A$_{v}$.

\subsection{Cosmic ray ionisation rate}\label{subsec:CR}

\begin{figure*}
     \centering
     \includegraphics[width=12cm]{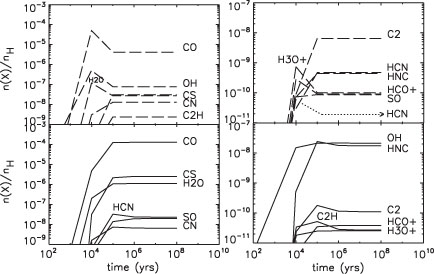}
     \caption{Influence of cosmic ray ionisation rate variations on
     some molecular abundances. We plot in a logarithmic scale the
     relative abundances n(X)/n$_{H}$ of various species, X, with
     respect to the time (in yrs). The upper two panels (dashed
     lines) are dedicated to Model 6 ($\zeta$ enhanced 100 times)
     while the lower two plots (black lines) show results for
     Model 0.}\label{fig:CR}
\end{figure*}

Using Table \ref{tab:trend_CR} and Fig. \ref{fig:CR}, we identify
five groups of species showing various degrees of sensitivity to
the change in the cosmic ray ionisation rate (from Model 0 with
$\zeta = 5 \times 10^{-17}$s$^{-1}$ to Model 6 with $\zeta = 5
\times 10^{-15}$s$^{-1}$, see Table \ref{tab:model_grid}). The
first group, including OCS, SO$_{2}$ or H$_{2}$CS, contains the
most sensitive species. Indeed these molecules have their
respective chemical abundances reduced by a factor of $\sim$
10$^{4}$ when $\zeta$ increases by two orders of magnitude. This
approximately quadratic response to the enhancement in $\zeta$
make them very good tracers of cosmic rays, especially in very
active star-forming regions or where an AGN contribution occurs.
However, this very high sensitivity leads certain species to
become undetectable at high $\zeta$. This is the case for OCS and
SO$_{2}$ whose relative abundances thus fall below our (arbitrary)
limit of detectability of $1 \times 10^{-12}$ for models with
$\zeta=$ 5 $\times 10^{-15}$s$^{-1}$.

A second category contains species such as SO, H$_{2}$S,
CH$_{2}$CO, CS, HNC or HCN whose abundances respond approximately
linearly to changes in $\zeta$. These species are good tracers of
cosmic rays as long as their fractional abundances are above the
detectability limit.

A third category of species includes molecules such as CO or
H$_{2}$O which show modest reduction of their chemical abundances
to increases in $\zeta$.

Species such as CN, OH or HCO$^{+}$ are fairly insensitive to
changes in $\zeta$.

Finally, we identify species whose fractional abundance increases
slightly (by a factor of $\sim$ 10; such as H$_{3}$O$^{+}$) or
more heavily (by a factor $\sim$ 100; such as C$_{2}$ and
C$_{2}$H) in response to an increase of $\zeta$. Ratios of
abundances of molecules from the first and fifth groups would be
highly sensitive tracers of $\zeta$.

The higher value of $\zeta$ used here generates loss rates for
most species that exceed other loss rates, and hence most
abundances decrease for higher $\zeta$. However, the larger
$\zeta$ also tends to increase the rate of forming ions such as
C$^{+}$, so that transient species such as C$_{2}$ and C$_{2}$H
increase.

\subsection{FUV radiation field}\label{subsec:FUV}

\begin{figure*}
     \centering
     \includegraphics[width=12cm]{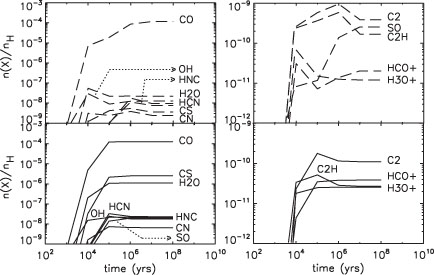}
     \caption{Influence of FUV radiation field variations on
     some molecular abundances. We plot in a logarithmic scale
     the relative abundances n(X)/n$_{H}$ of various species, X,
     with respect to the time (in yrs). The upper two panels
     (dashed lines) are dedicated to Model 7 (FUV radiation field
     enhanced 100 times) while the lower two plots (black lines)
     show results for Model 0.}\label{fig:FUV}
\end{figure*}

As expected, increasing the FUV radiation field by two orders of
magnitude leads to a general decrease of molecular abundances (see
Fig. \ref{fig:FUV} comparing Models 0 and 7 listed in Table
\ref{tab:model_grid}; see also Table \ref{tab:trend_FUV}).
Sulfur-bearing species appear also very sensitive to the changes
in the radiation field which makes them very good tracers of FUV
flux. Molecules such as H$_{2}$S or H$_{2}$CS show relative
abundances linearly reduced with respect to the FUV increase while
the CS, OCS and SO$_{2}$ molecules are more heavily reduced, by a
factor of $\sim$ 1000. However, SO is less sensitive. Molecules
CO, CN, OH, HNC, HCN, HCO$^{+}$ and H$_{3}$O$^{+}$ seem to be
rather insensitive to the changes in the FUV radiation field.

The small increase of the relative abundances of molecules such as
C$_{2}$ and C$_{2}$H is due to the injection of free carbon from
the photodestruction of carbon-bearing species. This injection
stimulates hydrocarbon growth.

\subsection{Density}\label{subsec:dens}

\begin{figure*}
     \centering
     \includegraphics[width=12cm]{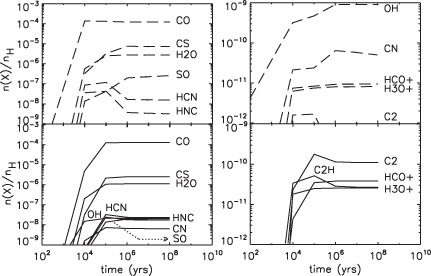}
     \caption{Influence of density variations on some molecular
     abundances. We plot in a logarithmic scale the relative
     abundances n(X)/n$_{H}$ of various species, X, with respect to
     the time (in yrs). The upper two panels (dashed lines) are dedicated
     to Model 8 (number density of $10^{6}$cm$^{-3}$) while the lower
     two plots (black lines) show results for Model 0.}\label{fig:dens}
\end{figure*}

We can roughly separate species into four main categories,
depending on their behavior when the density changes. Using
Fig.\ref{fig:dens} and Table \ref{tab:trend_dens}, we identify a
first category including species rather insensitive to gas density
changes. This includes molecules such as CO, H$_{2}$O, H$_{2}$CS,
CS, HCO$^{+}$, H$_{3}$O$^{+}$, HCN or HNC whose abundance
variations are below a factor of $\sim$5. This category is
followed by the molecules SO and OH which show moderate change in
their chemical abundances by a factor of $\approx$ 10 (especially
after 10$^{6}$yr, see Fig. \ref{fig:dens}). A third category
includes species that respond to an increase in gas number density
from 10$^{4}$ to 10$^{6}$ cm$^{-3}$ with heavy reduction (CN and
C$_{2}$, factors of 100), and very heavy reduction of their
fractional abundances (C$_{2}$H, CH$_{2}$CO or H$_{2}$CO, factors
from $>$100). These very heavy reductions usually lead to species
becoming non-detectable (see Table \ref{tab:trend_dens}). For this
reason, C$_{2}$H, CH$_{2}$CO H$_{2}$CO cannot be considered as
good tracers of density. On the contrary, CN and C$_{2}$ molecules
could be seen as very good tracers of gas density since they are
linearly linked with density changes. A last group of species able
to be recognized as very good tracers of density, contains the
molecules CO$_{2}$, OCS, H$_{2}$S and SO$_{2}$. The relative
abundances of these species indeed vary linearly (factor of 100)
with the change in gas density, except for the SO$_{2}$ which
varies quadratically with the gas density, promoting this molecule
to the best (detectable) tracer of n(H$_{2}$).

However these results take no account of the fact that freeze-out
of gas phase species has been omitted from these models. For gas
of number density n$_{H}$ and metallicity z, the freeze-out
timescale is $\approx
10^{6}$yr$(10^{4}/$n$_{H})($z$_{\odot}$/z$)$. Thus, at
n$_{H}=10^{4}$cm$^{-3}$ and z=z$_{\odot}$ the timescale is around
one million years, a plausible lifetime for gas in a PDR. However,
at n$_{H}=10^{6}$cm$^{-3}$, the timescale becomes unfeasibly short
unless the metallicity is very low. Thus, unless desorption
processes are operating efficiently \citep{Robe07} significant
freeze-out will be occurring in gas at very high densities.

\section{Predicted PDR tracers for selected galaxy types}\label{sec:gal}

In this section, we have selected several fairly well-known
galaxies as illustrating examples of the starburst, SB+AGN and
high-z galaxies. These galaxies are compared with our selection of
models in an attempt to identify, for each case, the best
molecular tracers of dense PDR gas. This study is of particular
interest for the preparations of the future observational programs
of Herschel and ALMA.

An interesting type of galaxy concerns objects such as M~83,
NGC~253, i.e. typical starbursts. It is commonly accepted that one
may reproduce such environments by using a FUV radiation field
enhanced as compared to normal spiral galaxies
\citep{Isra01,Baye06,Baye08a}. M~83 and NGC~253 show solar
metallicity (see \citealt{Orig04, Smit06} and \citealt{Zari94},
respectively), leading to a choice of standard elemental initial
abundance ratios and density. Model 7 appears consequently a good
model likely to represent this morphological type of galaxy.

Another morphological type of galaxy that we have investigated
concerns galaxies hosting an additional source such as active
nucleus (SB+AGN). Objects such as the galaxies Arp~220
\citep{Soif84, Joy86} or NGC 3079 \citep{Ford86, Sosa01} are
well-known examples of this category. Indeed, whether Arp~220 has
an AGN or not is still a very highly debated topic (see for
instance \citealt{Down07,Saka08,Aalt08}). Although NGC~3079
certainly contains an AGN, it seems that also for this source the
more extended SB may dominate the molecular gas properties (see
for instance \citealt{Kohn07}). Following the parameter choices
from \citet{Baye08a}, we have adopted here Model 6 to represent
these environments. Indeed, in this model, the cosmic ray
ionisation rate is enhanced 100 times ($\zeta = 5 \times
10^{-15}$s$^{-1}$) as compared to the standard values while the
FUV radiation has been maintained at $10^{3}$ I$_{\odot}$. The
selected metallicity is solar. The density, the initial elemental
abundance ratios and the A$_{v}$ values are standard. The model
adopted here is thus an attempt to simulate qualitatively XDR-like
environments. To reproduce ''pure" AGN sources, full XDR models
\citep{Meij05a,Meij05b,Spaa05} are however required.

Finally, we want to investigate galaxies at high redshift. Typical
examples to be considered are the Cloverleaf QSO and APM 08279
with their respective redshifts of z$_{red.} \approx$ 2.6 and
z$_{red.} \approx$ 3.9. For these sources, the physical properties
are unknown or highly uncertain, even if several molecules have
already been detected \citep{Garc06,Guel07,Krip07,Weis07,Riec09}.
The presence of both dense and less dense gas make these sources
particularly good illustrating examples for our study. Some of
these measurements \citep{Gao04b, Wu05, Wagg05} suggest that star
formation may be active, even if these galaxies are affected by an
AGN. \citet{Garc06} have suggested that the unusual chemistry
observed may be due to a combination of star formation and AGN
(PDR/XDR) chemistries. In an attempt to describe such sources, we
have run an additional model presented in Table
\ref{tab:model_grid} (Model 12). This model shows a high value of
the cosmic ray ionization rate, a high value of the radiation
field intensity, and a relatively low metallicity (all with
respect to the Milky Way).

We summarize in Table \ref{tab:gal} the species predicted to be
the best molecular tracers of dense PDR gas in each case (above
the limit of detectability arbitrarily taken to be equivalent to a
fractional abundance of 1$\times 10^{-12}$). Here our goal was to
give the reader some observational clues derived from this
modelling work. Our choice of galaxies are simply examples; none
of the models we are presenting in Table \ref{tab:gal} are meant
as detailed studies of own individual galaxy. Together with the
overall trends listed in Sect. \ref{sec:sensi}, Table
\ref{tab:gal} should enable a more detailed description of the
physical properties of the galaxy to be made.

Whichever model (6, 7 or 12) is used, CO and H$_{2}$O molecules
are predicted to have high fractional abundances (from $\approx
3\times 10^{-8}$ up to $10^{-4}$). More surprisingly, for Models 6
and 7, CS, CN and OH also show relatively high fractional
abundances, ranging between $10^{-9}$ and $10^{-8}$. According to
the estimates of \citet{Lint05} these abundances are sufficiently
large that unresolved active galaxies should be detectable in
these species. Observational tests of the predictions listed in
Table \ref{tab:gal} have been already performed by
\citet{Baye08b}. They have revealed the importance of CS lines in
starburst galaxies. Further observational tests of predictions
made in Table \ref{tab:gal} will be detailed in forthcoming
papers. For Model 12, CN and CS have chemical abundances which
drop severely by at least two orders of magnitude while the OH
relative abundance increases by a factor of 10 relative to Model
0. This feature again reinforces the interest of detecting these
species, especially at high redshift.

Finally, it is interesting to note that molecules such as HCN, HNC
and HCO$^{+}$ do not show especially high relative abundances in
the models investigated in this section. Indeed, the HCN and HNC
chemical abundances for Models 6, 7 and 12 fall to $\approx
10^{-11}$. Moreover, the HCO$^{+}$ abundance varies from 2.0
$\times 10^{-11}$ (Model 7) to 4.0 $\times 10^{-10}$ (Model 12).
However, we note that using a full XDR model \citep{Meij05b,
Spaa05, Meij07} might change the results of Model 6. Indeed, in a
full XDR model the cosmic rays considered are much more energetic
than the ones normally used in PDR models.

\section{Discussion and conclusions}\label{sec:conclu}

The main results of this study are the trends in the chemistry in
response to changes in the physical conditions (model parameters)
summarized in Tables \ref{tab:trend_meta} to \ref{tab:trend_dens}.
In addition to the evolution of the chemistry, Table \ref{tab:gal}
may be important in the context of the preparation of future
observational programs (Herschel, ALMA). Our results focus mainly
on the chemistry taking place at A$_{v} \approx 8$, i.e. in the
densest part of the PDR. Analysis and discussion on the chemistry
taking place in the outer region of the cloud (A$_{v} \ll 3$)
where the gas temperature and the FUV radiation field are higher
is beyond the scope of this paper. While previous studies
\citep{Isra95,Geri00,Isra01,Isra02,Isra03,Baye04,Baye06} focused
their attention on determining the physical conditions of a few
extragalactic star-forming or active regions using various PDR
codes and the emissions of only a couple of molecules (mainly
neutral carbon and carbon monoxide), we provide here a wide and
complete analysis of the chemistry, and report in details on 19
species in a large range of parameter space.

Our model predictions must be compared with observations (e.g.
\citealt{Radf91, Solo92, Solo04, Mart05, Evan06, Grac06, Mart06,
Aalt07c, Iman07, Knud07, Papa07}) to be validated. Recent
extragalactic detections of HCN and HCO$^{+}$ in 12 nearby
galaxies summarized in \citet{Krip08} provide useful estimations
of molecular abundances. When the source hosts an AGN
contribution, the relative HCN abundance they predicted from an
LVG analysis of their data, is in agreement with our predictions
(model 6), to within a factor of 10. For starburst-dominated
galaxies, we also obtain a good agreement. As another example,
H$_{3}$O$^{+}$ has been recently detected in Arp~220
\citep{VanDerTak08}. The H$_{3}$O$^{+}$ abundance in this SB+AGN
source is estimated to be 2-10$\times 10^{-9}$. Model 6 predicts
an abundance of H$_{3}$O$^{+}$ lower by a factor of 10. However,
\citet{VanDerTak08} mentioned that their abundance estimation is
somewhat hampered by source size uncertainties, which we think may
explain the discrepancy between our predictions and their
observations. We conclude that our models appear to be in harmony
with the limited observational data.

It is also particularly useful to compare the molecular tracers
predicted by dense star-forming core models \citep{Baye08a} with
those presented in this article (derived from the UCL\_PDR code).
Indeed, it is essential for better understanding and diagnosing
the nuclear energy source and activity in galaxies to be able to
disentangle the PDR emission from that produced by dense
star-forming cores. This should lead to a better determination of
the excitation conditions in such regions. These studies are of
importance because they concern not only the local Universe
\citep{Kohn01,Grac06,Baan08,Krip08} but also could help us to
better understand the galaxy formation in the Universe at higher
redshifts \citep{Ivis05a, Ivis05b, Ivis08, Youn08a, Youn08b}.
Comparing model predictions we thus have obtained that for the
starburst galaxy case, SO$_{2}$ and H$_{2}$S are predicted to be
likely inappropriate tracers of PDR regions while they are
predicted as good tracers of dense star-forming cores. By
contrast, HCO$^{+}$ is predicted to be undetectable in dense
star-forming cores while enhanced in PDR-dominated galaxies. For
high redshift sources, H$_{2}$CS and H$_{2}$CO are both predicted
to be inadequate molecular tracers of PDRs while both are good
signatures of dense star-forming cores.

By comparing the model predictions for starburst galaxies
\citep{Baye08a} and for the present study of PDR chemistry, we
note that while the molecular component of AGN-excited galaxies is
typical of PDRs, galaxies dominated by clusters of massive stars
(SB) contain molecular tracers of both PDR and dense star-forming
gas. Thus, in principle, it should be possible to distinguish
between SB+AGN and starburst excitations.

In summary, we have explored PDR chemistry for a range of
parameters representing galaxies with intense FUV or cosmic ray
sources. Our main results demonstrate the sensitivity of the
chemistry to the local physical conditions, and show that
molecular observations can be used to determine those conditions.
We have adopted very crude estimations of the relevant physical
conditions in several galaxy types and use our models to predict
molecular species that may be the best tracers of those
conditions.

\begin{table*}
    \caption{Input parameters of the UCL\_PDR models used to
    perform this study (see Sects. \ref{sec:para} and \ref{sec:sensi}).
    The abbreviation ``ST'' represents the standard values while
    the abbreviations ``CL02'', ``HW02'' and ``UN02'' are the initial
    elemental abundance ratios references detailed in Table
    \ref{tab:ele_ab_ratios}. This table does not present all the
    input parameters of the UCL\_PDR code for each model but only
    lists the parameters set to values different from the standard
    ones.}\label{tab:model_grid}
    \begin{center}
    \begin{tabular}{r c c c c c c c c}
    \hline
    Model & Metallicity   & Gas-to-dust              & H$_{2}$ form.            & A$_{v}$ & Ini. Elem.    & FUV rad. field & $\zeta$           & Gas Density\\
          & (z$_{\odot}$) & mass ratio (d$_{\odot}$) & rate coeff.(R$_{\odot}$) &         & Abund. ratios & (I$_{\odot}$)  & ($\zeta_{\odot}$) & (cm$^{-3}$)\\
    \hline
    0     & 1             & 1                        & 1                        & 8       & ST            & 10$^{3}$       & 1                 & 10$^{4}$\\
    1     & 1/2           & 1/2                      & 1/2                      & 8       & ST/2          & 10$^{3}$       & 1                 & 10$^{4}$\\
    2     & 1/4           & 1/4                      & 1/4                      & 8       & ST/4          & 10$^{3}$       & 1                 & 10$^{4}$\\
    3     & 1/10          & 1/10                     & 1/10                     & 8       & ST/10         & 10$^{3}$       & 1                 & 10$^{4}$\\
    4     & 1/100         & 1/100                    & 1/100                    & 8       & ST/100        & 10$^{3}$       & 1                 & 10$^{4}$\\
    5     & 1             & 1                        & 1                        & 3       & ST            & 10$^{3}$       & 1                 & 10$^{4}$\\
    6     & 1             & 1                        & 1                        & 8       & ST            & 10$^{3}$       & 100               & 10$^{4}$\\
    7     & 1             & 1                        & 1                        & 8       & ST            & 10$^{5}$       & 1                 & 10$^{4}$\\
    8     & 1             & 1                        & 1                        & 8       & ST            & 10$^{3}$       & 1                 & 10$^{6}$\\
    9     & 1             & 1                        & 1                        & 8       & CL02          & 10$^{3}$       & 1                 & 10$^{4}$\\
   10     & 1             & 1                        & 1                        & 8       & HW02          & 10$^{3}$       & 1                 & 10$^{4}$\\
   11     & 1             & 1                        & 1                        & 8       & UN02          & 10$^{3}$       & 1                 & 10$^{4}$\\
   12     & 1/10          & 1/10                     & 1/10                     & 8       & ST/10         & 10$^{5}$       & 100               & 10$^{4}$\\
   \hline
    \end{tabular}
    \end{center}
\end{table*}

\begin{table*}
    \caption{Initial abundance ratios used in Table \ref{tab:model_grid}.
    The abbreviations ``CL02'', ``HW02'' and ``UN02'' refer to
    \citet{Chie02}, \citet{Hege02} and \citet{Umed02}, respectively (see
    Sect. \ref{sec:para}). The standard initial elemental abundance ratios
    (``ST'') are from \citet{Sava96, Sofi97, Meye98, Snow02, Knau03}. Models
    using the ``ST'' values of the initial elemental abundance ratios are
    the models corresponding to z = z$_{\odot}$.}\label{tab:ele_ab_ratios}
    \begin{center}
    \begin{tabular}{c c c c c}
    \hline
    & ST & CL02 & HW02 & UN02 \\
    \hline
    C/H & 1.4$\times$10$^{-4}$ & 1.4$\times$10$^{-4}$ & 1.4$\times$10$^{-4}$ &
    1.4$\times$10$^{-4}$\\
    O/H & 3.2$\times$10$^{-4}$ & 4.54$\times$10$^{-4}$ & 1.53$\times$10$^{-3}$ &
    1.18$\times$10$^{-3}$\\
    N/H & 6.5$\times$10$^{-5}$ & 5.99$\times$10$^{-11}$ & 1.58$\times$10$^{-9}$ &
    3.24$\times$10$^{-7}$\\
    S/H & 1.4$\times$10$^{-6}$ & 7.59$\times$10$^{-6}$ & 1.54$\times$10$^{-4}$ &
    4.06$\times$10$^{-5}$\\
    He/H & 7.5$\times$10$^{-2}$ & 7.5$\times$10$^{-2}$ & 7.5$\times$10$^{-2}$ &
    7.5$\times$10$^{-2}$\\
    Mg/H & 5.1$\times$10$^{-6}$ & 1.83$\times$10$^{-5}$ & 1.2$\times$10$^{-4}$ &
    6.18$\times$10$^{-5}$\\
    \hline
    \end{tabular}
    \end{center}
\end{table*}

\begin{table*}
    \caption{Trends of molecular fractional abundances, computed for
    different values of metallicity. When a single species is defined as
    ``undetectable'' or ``not detectable'', it means that its relative
    abundance is below the adopted limit of detectability of
    $1 \times 10^{-12}$. }\label{tab:trend_meta}
    \begin{center}
    \begin{tabular}{l l}
    \hline
    Molecule & Response to metallicity changes\\
    \hline
    CO, H$_{2}$O & linear tracers of metallicity\\
    CS, SO & for z$>$1/100z$_{\odot}$, good linear tracers of metallicity\\
    CN & for z$>$1/100z$_{\odot}$; insensitive to metallicity changes\\
    OH, H$_{3}$O$^{+}$ & most abundant at z=1/100z$_{\odot}$; generally insensitive to z\\
    HNC, HCN & for z$>$1/10z$_{\odot}$, insensitive to metallicity changes\\
    HCO$^{+}$ & always above the limit of detectability; insensitive to metallicity changes\\
    C$_{2}$, C$_{2}$H & for z$>$1/10z$_{\odot}$, inversely dependent on z changes\\
    \hline
    CO$_{2}$, OCS & good linear tracers of metallicity except for z$<$1/100z$_{\odot}$ (undetectable)\\
    SO$_{2}$,H$_{2}$S & for z$>$1/100z$_{\odot}$, good linear tracer of metallicity\\
    H$_{2}$CS, CH$_{2}$CO & for z$>$1/10z$_{\odot}$, inversely dependent on z changes\\
    H$_{2}$CO & for z$>$1/100z$_{\odot}$, inversely dependent on z changes\\
    \hline
    \end{tabular}
    \end{center}
\end{table*}

\begin{table*}
    \caption{Trends of molecular fractional abundances, computed for different values of the initial elemental abundance ratios.
    Comparison of the PDR chemistry for various values of initial abundances (see Table \ref{tab:ele_ab_ratios} and Models
    9, 10 \& 11 in Table \ref{tab:model_grid}) while other parameters have the
    standard values (see Table \ref{tab:stand}).}\label{tab:trend_ele_ab_ratios}
    \begin{center}
    \begin{tabular}{l l}
    \hline
    Molecule & Response to initial elemental abundance ratios changes\\
    \hline
    CO, H$_{2}$O & insensitive to changes of any initial elemental abundance ratios\\
    CS & sensitive to changes in both S/H and O/H; variations within a factor of $\geq$ 50\\
    SO & very good tracer of S/H; variations up to a factor $\approx$ 1000; very sensitive to O/H changes\\
    CN & highest abundance for N-rich models; very good tracer of N/H\\
    OH & insensitive to any changes in initial elemental abundance ratios\\
    HNC, HCN & very good tracers of N/H\\
    HCO$^{+}$, H$_{3}$O$^{+}$ & poor tracer of initial elemental abundance ratios\\
    C$_{2}$, C$_{2}$H & surviving only if N-rich environments\\
    \hline
    CO$_{2}$ & insensitive in changes of any initial elemental abundance ratios\\
    OCS & highest abundance for S- and O-rich models; variations within a factor of $\approx$ 100; good tracer\\
    SO$_{2}$ & very good tracer of S/H and O/H\\
    H$_{2}$S & insensitive to any changes in initial elemental abundance ratios\\
    H$_{2}$CS & bad tracer of S/H and O/H; lowest abundance for the S-and O-rich models\\
    CH$_{2}$CO, H$_{2}$CO & abundant when N-enriched environment, poor tracers of S/H and O/H\\
    \hline
    \end{tabular}
    \end{center}
\end{table*}

\begin{table*}
    \caption{Trends of molecular fractional abundances, computed for two values
    of the opacity (A$_{v}$). Comparison of the PDR chemistry for
    A$_{v}$ = 8 (Model 0) and A$_{v}$ = 3 (Model 5) while other parameters have the
    standard values (see Table \ref{tab:stand}).}\label{tab:trend_Av}
    \begin{center}
    \begin{tabular}{l l}
    \hline
    Molecule & Response to opacity changes\\
    \hline
    CO & insensitive to opacity changes\\
    H$_{2}$O, HNC, HCN, SO & very good tracers of A$_{v}$\\
    CS & at A$_{v}$ = 3, very good tracer of the opacity\\
    CN, HCO$^{+}$, H$_{3}$O$^{+}$ & not very sensitive to opacity changes\\
    OH & insensitive to opacity changes; poor tracer of A$_{v}$\\
    C$_{2}$, C$_{2}$H & insensitive to opacity changes\\
    \hline
    CO$_{2}$ & at A$_{v}$ = 3, abundance decreases by a factor of 10$^{4}$; very good tracer of A$_{v}$\\
    OCS, H$_{2}$CS & at A$_{v}$ = 3, reduced by a factor of 10$^{4}$; become undetectable\\
    SO$_{2}$ & at A$_{v}$ = 3, reduced by a factor of 10$^{6}$; become undetectable\\
    H$_{2}$S, CH$_{2}$CO & at A$_{v}$ = 3, decreased by a factor $\ge$10$^{2}$; become undetectable\\
    H$_{2}$CO & when A$_{v}$ = 3, abundance reduction by a factor of 10; become undetectable\\
    \hline
    \end{tabular}
    \end{center}
\end{table*}

\begin{table*}
    \caption{Trends of molecular fractional abundances, computed for different values of the cosmic ray
    ionization rate ($\zeta$). Comparison of the PDR chemistry for
    $\zeta$ = 1$\zeta_{\odot}$ = 5 $\times 10^{-17}$s$^{-1}$ (Model 0) and $\zeta$ = 100$\times \zeta_{\odot}$ =
    5 $\times 10^{-15}$s$^{-1}$ (Model 6) while other parameters have the
    standard values (see Table \ref{tab:stand}).}\label{tab:trend_CR}
    \begin{center}
    \begin{tabular}{l l}
    \hline
    Molecule & Response to cosmic ray ionisation rate changes\\
    \hline
    CO, H$_{2}$O & reduced by a factor of 10 with $\zeta$; possible tracers of $\zeta$\\
    CS, SO, HNC, HCN & heavy (linear) abundance reduction by factor of 100 with $\zeta$; good tracers of $\zeta$ \\
    CN, OH, HCO$^{+}$ & insensitive to $\zeta$\\
    H$_{3}$O$^{+}$ & slight increase by factor of 10 with $\zeta$\\
    C$_{2}$, C$_{2}$H & heavy (linear) abundance increase by a factor of 100 with $\zeta$\\
    \hline
    CO$_{2}$, H$_{2}$CS & very heavy reduction by a factor of $\sim$ 10$^{4}$;  very good tracer of $\zeta$\\
    OCS, SO$_{2}$ & very heavily reduced by factor of $\sim$ 10$^{4}$ with $\zeta$; become undetectable\\
    H$_{2}$S, CH$_{2}$CO & heavy (linear) abundance reduction by factor of 100 with $\zeta$; become undetectable\\
    H$_{2}$CO & insensitive to $\zeta$\\
    \hline
    \end{tabular}
    \end{center}
\end{table*}

\begin{table*}
    \caption{Trends of molecular fractional abundances, computed for different values of the FUV radiation field.
    Comparison of the PDR chemistry for $I = 10^{3}$ I$_{\odot}$ (Model 0) and I$^{'} = 100 \times$ I$ = 10^{5} \times$ I$_{\odot}$ (Model 7)
    while other parameters have the
    standard values (see Table \ref{tab:stand}).}\label{tab:trend_FUV}
    \begin{center}
    \begin{tabular}{l l}
    \hline
    Molecule & Response to FUV radiation field changes\\
    \hline
    CO & insensitive to FUV changes\\
    H$_{2}$O & reduced by a factor of 100 (linearly to the FUV change); good tracer\\
    CS & heavy abundance reduction by a factor of 1000 with FUV; very sensitive to FUV changes\\
    SO & moderately reduced by a factor of 10 with FUV; possible tracer of FUV\\
    CN, OH, HNC,  & insensitive to the changes in FUV\\
    HCN, HCO$^{+}$, H$_{3}$O$^{+}$ & insensitive to the changes in FUV\\
    C$_{2}$, C$_{2}$H & slight abundance increase by a factor $\le$ 10 with FUV changes\\
    \hline
    CO$_{2}$, H$_{2}$S, H$_{2}$CS & abundance decreases (linearly) by a factor of 100 with the FUV changes; good tracers\\
    OCS, SO$_{2}$ & heavy reduction by a factor of 1000; very sensitive to FUV changes; very good tracer\\
    CH$_{2}$CO, H$_{2}$CO & insensitive to FUV changes\\
    \hline
    \end{tabular}
    \end{center}
\end{table*}

\begin{table*}
    \caption{Trends of molecular fractional abundances, computed for different values of the density.
    Comparison of the PDR chemistry for n(H$_{2}$)= 10$^{3}$ cm$^{-3}$ (Model 0) and n(H$_{2}$)= 10$^{6}$ cm$^{-3}$ (Model 6) while other parameters have the
    standard values (see Table \ref{tab:stand}).}\label{tab:trend_dens}
    \begin{center}
    \begin{tabular}{l l}
    \hline
    Molecule & Response to density changes\\
    \hline
    CO, H$_{2}$O, CS, HNC, HCN & abundance variations within a factor of $\le$ 4; poor tracers \\
    SO & abundance increase by a factor of $\sim$10 with density; possible tracer of density\\
    CN, C$_{2}$ & decrease by a factor of 100 (linearly inverse with the change of density); very good tracer\\
    OH & decrease by a factor of $\sim$10; possible tracer of density\\
    HCO$^{+}$, H$_{3}$O$^{+}$ & abundance variations within a factor of $\le$ 4; insensitive to density changes\\
    C$_{2}$H & decrease by a factor of $\ge$100; undetectable at high gas density\\
    \hline
    CO$_{2}$, OCS, H$_{2}$S & increase by a factor of $\ge$100; very good (linear) tracers of n(H$_{2}$)\\
    SO$_{2}$ & increase by a factor of $\ge$1000; very good (quadratic) tracer of n(H$_{2}$)\\
    H$_{2}$CS & abundance variations within a factor of $\sim$ 2; insensitive to density changes\\
    CH$_{2}$CO & decrease by a factor of $\ge$10$^{5}$; undetectable at high density\\
    H$_{2}$CO & (linear) decrease by a factor of $\ge$100; undetectable at n(H$_{2}$)=10$^{6}$cm$^{-3}$\\
    \hline
    \end{tabular}
    \end{center}
\end{table*}

\begin{table}
    \caption{Detectability of 19 species likely to
    trace the dense PDR gas component in three categories
    of galaxies (see Sect. \ref{sec:gal}). The limit of
    detectability has been taken to be (n(X)/n$_{H}$) = 1 $\times 10^{-12}$,
    as is typical for molecules in the Milky Way. Under this
    limit the species are not detectable (symbol $-$).
    Otherwise, they are marked with the symbol $+$. We
    have separated the species plotted in Figs. \ref{fig:met}
    to \ref{fig:ele_ab_ratios} from those only listed in
    Tables \ref{tab:trend_meta} to
    \ref{tab:trend_ele_ab_ratios}.}\label{tab:gal}
    \begin{center}
    \begin{tabular}{c c  c c}
    \hline
    & Model 7 & Model 6 & Model 12\\
    \hline
    Galaxy type  &  Starburst & SB+AGN & high redshift \\
    \hline
    CO & $+$ & $+$ & $+$\\
    H$_{2}$O & $+$ & $+$ & $+$\\
    CS & $+$ & $+$ & $+$\\
    SO & $+$ & $+$ & $+$\\
    CN & $+$ & $+$ & $+$\\
    OH & $+$ & $+$ & $+$\\
    HNC & $+$ & $+$ & $+$\\
    HCN & $+$ & $+$ & $+$\\
    HCO$^{+}$  & $+$ & $+$ & $+$\\
    H$_{3}$O$^{+}$  & $+$ & $+$ & $+$\\
    C$_{2}$ & $+$ & $+$ & $+$\\
    C$_{2}$H & $+$ & $+$ & $+$\\
    \hline
    CO$_{2}$  & $+$ & $+$ & $+$\\
    OCS  & $+$ & $-$ & $-$ \\
    SO$_2$  & $-$ & $-$ & $-$ \\
    H$_2$S  & $-$ & $-$ & $-$ \\
    H$_2$CS & $+$ & $+$ & $-$\\
    H$_2$CO & $+$ & $+$  & $-$\\
    CH$_2$CO & $+$ & $-$ & $-$\\
    \hline
    \end{tabular}
    \end{center}
\end{table}

\acknowledgments
\begin{acknowledgements}
    Acknowledgments

EB acknowledges financial support from the Leverhulme Trust.
\end{acknowledgements}

\bibliographystyle{apj}

\bibliography{references}

\begin{thebibliography}{73}
\expandafter\ifx\csname natexlab\endcsname\relax\def\natexlab#1{#1}\fi

\bibitem[{{Aalto} {et~al.}(2007){Aalto}, {Monje}, \& {Mart{\'{\i}}n}}]{Aalt07c}
{Aalto}, S., {Monje}, R., \& {Mart{\'{\i}}n}, S. 2007, \aap, 475, 479

\bibitem[{{Aalto} {et~al.}(2008){Aalto}, {Wilner}, {Spaans}, {Wiedner},
  {Sakamoto}, {Black}, \& {Caldas}}]{Aalt08}
{Aalto}, S., {Wilner}, D., {Spaans}, M., {Wiedner}, M.~C., {Sakamoto}, K.,
  {Black}, J.~H., \& {Caldas}, M. 2008, ArXiv e-prints

\bibitem[{{Baan} {et~al.}(2008){Baan}, {Henkel}, {Loenen}, {Baudry}, \&
  {Wiklind}}]{Baan08}
{Baan}, W.~A., {Henkel}, C., {Loenen}, A.~F., {Baudry}, A., \& {Wiklind}, T.
  2008, \aap, 477, 747

\bibitem[{{Bayet} {et~al.}(2004){Bayet}, {Gerin}, {Phillips}, \&
  {Contursi}}]{Baye04}
{Bayet}, E., {Gerin}, M., {Phillips}, T.~G., \& {Contursi}, A. 2004, \aap, 427,
  45

\bibitem[{{Bayet} {et~al.}(2006){Bayet}, {Gerin}, {Phillips}, \&
  {Contursi}}]{Baye06}
---. 2006, \aap, 460, 467

\bibitem[{{Bayet} {et~al.}(2008{\natexlab{a}}){Bayet}, {Lintott}, {Viti},
  {Mart{\'{\i}}n-Pintado}, {Mart{\'{\i}}n}, {Williams}, \&
  {Rawlings}}]{Baye08b}
{Bayet}, E., {Lintott}, C., {Viti}, S., {Mart{\'{\i}}n-Pintado}, J.,
  {Mart{\'{\i}}n}, S., {Williams}, D.~A., \& {Rawlings}, J.~M.~C.
  2008{\natexlab{a}}, \apjl, 685, L35

\bibitem[{{Bayet} {et~al.}(2009){Bayet}, {Lintott}, {Viti},
  {Mart{\'{\i}}n-Pintado}, {Mart{\'{\i}}n}, {Williams}, \&
  {Rawlings}}]{Baye08c}
---. 2009, \mnras, in preparation

\bibitem[{{Bayet} {et~al.}(2008{\natexlab{b}}){Bayet}, {Viti}, {Williams}, \&
  {Rawlings}}]{Baye08a}
{Bayet}, E., {Viti}, S., {Williams}, D.~A., \& {Rawlings}, J.~M.~C.
  2008{\natexlab{b}}, \apj, 676, 978

\bibitem[{{Bell} {et~al.}(2006){Bell}, {Roueff}, {Viti}, \&
  {Williams}}]{Bell06}
{Bell}, T.~A., {Roueff}, E., {Viti}, S., \& {Williams}, D.~A. 2006, \mnras,
  371, 1865

\bibitem[{{Bell} {et~al.}(2007){Bell}, {Viti}, \& {Williams}}]{Bell07}
{Bell}, T.~A., {Viti}, S., \& {Williams}, D.~A. 2007, \mnras, 378, 983

\bibitem[{{Boreiko} \& {Betz}(1993)}]{Bore93}
{Boreiko}, R.~T. \& {Betz}, A.~L. 1993, \apjl, 405, L39

\bibitem[{{Chieffi} \& {Limongi}(2002)}]{Chie02}
{Chieffi}, A. \& {Limongi}, M. 2002, \apj, 577, 281

\bibitem[{{Downes} \& {Eckart}(2007)}]{Down07}
{Downes}, D. \& {Eckart}, A. 2007, \aap, 468, L57

\bibitem[{{Evans} {et~al.}(2006){Evans}, {Solomon}, {Tacconi}, {Vavilkin}, \&
  {Downes}}]{Evan06}
{Evans}, A.~S., {Solomon}, P.~M., {Tacconi}, L.~J., {Vavilkin}, T., \&
  {Downes}, D. 2006, \aj, 132, 2398

\bibitem[{{Ford} {et~al.}(1986){Ford}, {Dahari}, {Jacoby}, {Crane}, \&
  {Ciardullo}}]{Ford86}
{Ford}, H.~C., {Dahari}, O., {Jacoby}, G.~H., {Crane}, P.~C., \& {Ciardullo},
  R. 1986, \apjl, 311, L7

\bibitem[{{Gao} \& {Solomon}(2004)}]{Gao04b}
{Gao}, Y. \& {Solomon}, P.~M. 2004, \apjs, 152, 63

\bibitem[{{Garc{\'{\i}}a-Burillo} {et~al.}(2006){Garc{\'{\i}}a-Burillo},
  {Graci{\'a}-Carpio}, {Gu{\'e}lin}, {Neri}, {Cox}, {Planesas}, {Solomon},
  {Tacconi}, \& {Vanden Bout}}]{Garc06}
{Garc{\'{\i}}a-Burillo}, S., {Graci{\'a}-Carpio}, J., {Gu{\'e}lin}, M., {Neri},
  R., {Cox}, P., {Planesas}, P., {Solomon}, P.~M., {Tacconi}, L.~J., \& {Vanden
  Bout}, P.~A. 2006, \apjl, 645, L17

\bibitem[{{Gerin} \& {Phillips}(2000)}]{Geri00}
{Gerin}, M. \& {Phillips}, T.~G. 2000, \apj, 537, 644

\bibitem[{{Gerin} \& {Phillips}(2001)}]{Geri01}
---. 2001, Astrophysics and Space Science Supplement, 277, 75

\bibitem[{{Graci{\'a}-Carpio} {et~al.}(2006){Graci{\'a}-Carpio},
  {Garc{\'{\i}}a-Burillo}, {Planesas}, \& {Colina}}]{Grac06}
{Graci{\'a}-Carpio}, J., {Garc{\'{\i}}a-Burillo}, S., {Planesas}, P., \&
  {Colina}, L. 2006, \apjl, 640, L135

\bibitem[{{Gu{\'e}lin} {et~al.}(2007){Gu{\'e}lin}, {Salom{\'e}}, {Neri},
  {Garc{\'{\i}}a-Burillo}, {Graci{\'a}-Carpio}, {Cernicharo}, {Cox},
  {Planesas}, {Solomon}, {Tacconi}, \& {vanden Bout}}]{Guel07}
{Gu{\'e}lin}, M., {Salom{\'e}}, P., {Neri}, R., {Garc{\'{\i}}a-Burillo}, S.,
  {Graci{\'a}-Carpio}, J., {Cernicharo}, J., {Cox}, P., {Planesas}, P.,
  {Solomon}, P.~M., {Tacconi}, L.~J., \& {vanden Bout}, P. 2007, \aap, 462, L45

\bibitem[{{Habing}(1968)}]{Habi68}
{Habing}, H.~J. 1968, \bain, 19, 421

\bibitem[{{Heger} \& {Woosley}(2002)}]{Hege02}
{Heger}, A. \& {Woosley}, S.~E. 2002, \apj, 567, 532

\bibitem[{{Imanishi} {et~al.}(2007){Imanishi}, {Nakanishi}, {Tamura}, {Oi}, \&
  {Kohno}}]{Iman07}
{Imanishi}, M., {Nakanishi}, K., {Tamura}, Y., {Oi}, N., \& {Kohno}, K. 2007,
  \aj, 134, 2366

\bibitem[{{Israel} \& {Baas}(2001)}]{Isra01}
{Israel}, F.~P. \& {Baas}, F. 2001, \aap, 371, 433

\bibitem[{{Israel} \& {Baas}(2002)}]{Isra02}
---. 2002, \aap, 383, 82

\bibitem[{{Israel} \& {Baas}(2003)}]{Isra03}
---. 2003, \aap, 404, 495

\bibitem[{{Israel} {et~al.}(1995){Israel}, {White}, \& {Baas}}]{Isra95}
{Israel}, F.~P., {White}, G.~J., \& {Baas}, F. 1995, \aap, 302, 343

\bibitem[{{Ivison} {et~al.}(2008){Ivison}, {Morrison}, {Biggs}, {Smail},
  {Willner}, {Gurwell}, {Greve}, {Stevens}, \& {Ashby}}]{Ivis08}
{Ivison}, R.~J., {Morrison}, G.~E., {Biggs}, A.~D., {Smail}, I., {Willner},
  S.~P., {Gurwell}, M.~A., {Greve}, T.~R., {Stevens}, J.~A., \& {Ashby},
  M.~L.~N. 2008, \mnras, 390, 1117

\bibitem[{{Ivison} {et~al.}(2005{\natexlab{a}}){Ivison}, {Smail}, {Bentz},
  {Stevens}, {Men{\'e}ndez-Delmestre}, {Chapman}, \& {Blain}}]{Ivis05b}
{Ivison}, R.~J., {Smail}, I., {Bentz}, M., {Stevens}, J.~A.,
  {Men{\'e}ndez-Delmestre}, K., {Chapman}, S.~C., \& {Blain}, A.~W.
  2005{\natexlab{a}}, \mnras, 362, 535

\bibitem[{{Ivison} {et~al.}(2005{\natexlab{b}}){Ivison}, {Smail}, {Dunlop},
  {Greve}, {Swinbank}, {Stevens}, {Mortier}, {Serjeant}, {Targett}, {Bertoldi},
  {Blain}, \& {Chapman}}]{Ivis05a}
{Ivison}, R.~J., {Smail}, I., {Dunlop}, J.~S., {Greve}, T.~R., {Swinbank},
  A.~M., {Stevens}, J.~A., {Mortier}, A.~M.~J., {Serjeant}, S., {Targett},
  T.~A., {Bertoldi}, F., {Blain}, A.~W., \& {Chapman}, S.~C.
  2005{\natexlab{b}}, \mnras, 364, 1025

\bibitem[{{Joy} {et~al.}(1986){Joy}, {Lester}, {Harvey}, \& {Frueh}}]{Joy86}
{Joy}, M., {Lester}, D.~F., {Harvey}, P.~M., \& {Frueh}, M. 1986, \apj, 307,
  110

\bibitem[{{Knauth} {et~al.}(2003){Knauth}, {Andersson}, {McCandliss}, \&
  {Moos}}]{Knau03}
{Knauth}, D.~C., {Andersson}, B.-G., {McCandliss}, S.~R., \& {Moos}, H.~W.
  2003, \apjl, 596, L51

\bibitem[{{Knudsen} {et~al.}(2007){Knudsen}, {Walter}, {Weiss}, {Bolatto},
  {Riechers}, \& {Menten}}]{Knud07}
{Knudsen}, K.~K., {Walter}, F., {Weiss}, A., {Bolatto}, A., {Riechers}, D.~A.,
  \& {Menten}, K. 2007, \apj, 666, 156

\bibitem[{{Kohno} {et~al.}(2001){Kohno}, {Matsushita}, {Vila-Vilar{\' o}},
  {Okumura}, {Shibatsuka}, {Okiura}, {Ishizuki}, \& {Kawabe}}]{Kohn01}
{Kohno}, K., {Matsushita}, S., {Vila-Vilar{\' o}}, B., {Okumura}, S.~K.,
  {Shibatsuka}, T., {Okiura}, M., {Ishizuki}, S., \& {Kawabe}, R. 2001, in ASP
  Conf. Ser. 249: The Central Kiloparsec of Starbursts and AGN: The La Palma
  Connection, 672

\bibitem[{{Kohno} {et~al.}(2007){Kohno}, {Nakanishi}, \& {Imanishi}}]{Kohn07}
{Kohno}, K., {Nakanishi}, K., \& {Imanishi}, M. 2007, in Astronomical Society
  of the Pacific Conference Series, Vol. 373, The Central Engine of Active
  Galactic Nuclei, ed. L.~C. {Ho} \& J.-W. {Wang}, 647--+

\bibitem[{{Krips} {et~al.}(2008){Krips}, {Neri}, {Garc{\'{\i}}a-Burillo},
  {Mart{\'{\i}}n}, {Combes}, {Graci{\'a}-Carpio}, \& {Eckart}}]{Krip08}
{Krips}, M., {Neri}, R., {Garc{\'{\i}}a-Burillo}, S., {Mart{\'{\i}}n}, S.,
  {Combes}, F., {Graci{\'a}-Carpio}, J., \& {Eckart}, A. 2008, \apj, 677, 262

\bibitem[{{Krips} {et~al.}(2007){Krips}, {Peck}, {Sakamoto}, {Petitpas},
  {Wilner}, {Matsushita}, \& {Iono}}]{Krip07}
{Krips}, M., {Peck}, A.~B., {Sakamoto}, K., {Petitpas}, G.~B., {Wilner}, D.~J.,
  {Matsushita}, S., \& {Iono}, D. 2007, \apjl, 671, L5

\bibitem[{{Le Teuff} {et~al.}(2000){Le Teuff}, {Millar}, \&
  {Markwick}}]{LeTe00}
{Le Teuff}, Y.~H., {Millar}, T.~J., \& {Markwick}, A.~J. 2000, \aaps, 146, 157

\bibitem[{{Lintott} \& {Viti}(2006)}]{Lint06}
{Lintott}, C. \& {Viti}, S. 2006, \apjl, 646, L37

\bibitem[{{Lintott} {et~al.}(2005){Lintott}, {Viti}, {Williams}, {Rawlings}, \&
  {Ferreras}}]{Lint05}
{Lintott}, C.~J., {Viti}, S., {Williams}, D.~A., {Rawlings}, J.~M.~C., \&
  {Ferreras}, I. 2005, \mnras, 360, 1527

\bibitem[{{Mart{\'{\i}}n} {et~al.}(2005){Mart{\'{\i}}n},
  {Mart{\'{\i}}n-Pintado}, {Mauersberger}, {Henkel}, \&
  {Garc{\'{\i}}a-Burillo}}]{Mart05}
{Mart{\'{\i}}n}, S., {Mart{\'{\i}}n-Pintado}, J., {Mauersberger}, R., {Henkel},
  C., \& {Garc{\'{\i}}a-Burillo}, S. 2005, \apj, 620, 210

\bibitem[{{Mart{\'{\i}}n} {et~al.}(2006){Mart{\'{\i}}n}, {Mauersberger},
  {Mart{\'{\i}}n-Pintado}, {Henkel}, \& {Garc{\'{\i}}a-Burillo}}]{Mart06}
{Mart{\'{\i}}n}, S., {Mauersberger}, R., {Mart{\'{\i}}n-Pintado}, J., {Henkel},
  C., \& {Garc{\'{\i}}a-Burillo}, S. 2006, \apjs, 164, 450

\bibitem[{{Meijerink} \& {Spaans}(2005)}]{Meij05b}
{Meijerink}, R. \& {Spaans}, M. 2005, \aap, 436, 397

\bibitem[{{Meijerink} {et~al.}(2007){Meijerink}, {Spaans}, \&
  {Israel}}]{Meij07}
{Meijerink}, R., {Spaans}, M., \& {Israel}, F.~P. 2007, \aap, 461, 793

\bibitem[{{Meijerink} {et~al.}(2005){Meijerink}, {Tilanus}, {Dullemond},
  {Israel}, \& {van der Werf}}]{Meij05a}
{Meijerink}, R., {Tilanus}, R.~P.~J., {Dullemond}, C.~P., {Israel}, F.~P., \&
  {van der Werf}, P.~P. 2005, \aap, 430, 427

\bibitem[{{Meyer} {et~al.}(1998){Meyer}, {Jura}, \& {Cardelli}}]{Meye98}
{Meyer}, D.~M., {Jura}, M., \& {Cardelli}, J.~A. 1998, \apj, 493, 222

\bibitem[{{Origlia} {et~al.}(2004){Origlia}, {Ranalli}, {Comastri}, \&
  {Maiolino}}]{Orig04}
{Origlia}, L., {Ranalli}, P., {Comastri}, A., \& {Maiolino}, R. 2004, \apj,
  606, 862

\bibitem[{{Pagani} {et~al.}(2007){Pagani}, {Bacmann}, {Cabrit}, \&
  {Vastel}}]{Paga07}
{Pagani}, L., {Bacmann}, A., {Cabrit}, S., \& {Vastel}, C. 2007, \aap, 467, 179

\bibitem[{{Papadopoulos}(2007)}]{Papa07}
{Papadopoulos}, P.~P. 2007, \apj, 656, 792

\bibitem[{{Pickles} \& {Williams}(1981)}]{Pick81}
{Pickles}, J.~B. \& {Williams}, D.~A. 1981, \apss, 80, 337

\bibitem[{{Radford} {et~al.}(1991){Radford}, {Downes}, \& {Solomon}}]{Radf91}
{Radford}, S.~J.~E., {Downes}, D., \& {Solomon}, P.~M. 1991, \apjl, 368, L15

\bibitem[{{Riechers} {et~al.}(2009){Riechers}, {Walter}, {Carilli}, \&
  {Lewis}}]{Riec09}
{Riechers}, D.~A., {Walter}, F., {Carilli}, C.~L., \& {Lewis}, G.~F. 2009,
  \apj, 690, 463

\bibitem[{{Roberts} {et~al.}(2007){Roberts}, {Rawlings}, {Viti}, \&
  {Williams}}]{Robe07}
{Roberts}, J.~F., {Rawlings}, J.~M.~C., {Viti}, S., \& {Williams}, D.~A. 2007,
  \mnras, 382, 733

\bibitem[{{R{\"o}llig} {et~al.}(2007){R{\"o}llig}, {Abel}, {Bell}, {Bensch},
  {Black}, {Ferland}, {Jonkheid}, {Kamp}, {Kaufman}, {Le Bourlot}, {Le Petit},
  \& {etc, }}]{Roel07}
{R{\"o}llig}, M., {Abel}, N.~P., {Bell}, T., {Bensch}, F., {Black}, J.,
  {Ferland}, G.~J., {Jonkheid}, B., {Kamp}, I., {Kaufman}, M.~J., {Le Bourlot},
  J., {Le Petit}, F., \& {etc, }. 2007, \aap, 467, 187

\bibitem[{{Sakamoto} {et~al.}(2008){Sakamoto}, {Wang}, {Wiedner}, {Wang},
  {Peck}, {Zhang}, {Petitpas}, {Ho}, \& {Wilner}}]{Saka08}
{Sakamoto}, K., {Wang}, J., {Wiedner}, M.~C., {Wang}, Z., {Peck}, A.~B.,
  {Zhang}, Q., {Petitpas}, G.~R., {Ho}, P.~T.~P., \& {Wilner}, D.~J. 2008,
  \apj, 684, 957

\bibitem[{{Sembach} \& {Savage}(1996)}]{Sava96}
{Sembach}, K.~R. \& {Savage}, B.~D. 1996, \apj, 457, 211

\bibitem[{{Smith} {et~al.}(2006){Smith}, {Westmoquette}, {Gallagher},
  {O'Connell}, {Rosario}, \& {de Grijs}}]{Smit06}
{Smith}, L.~J., {Westmoquette}, M.~S., {Gallagher}, J.~S., {O'Connell}, R.~W.,
  {Rosario}, D.~J., \& {de Grijs}, R. 2006, \mnras, 370, 513

\bibitem[{{Snow} {et~al.}(2002){Snow}, {Rachford}, \& {Figoski}}]{Snow02}
{Snow}, T.~P., {Rachford}, B.~L., \& {Figoski}, L. 2002, \apj, 573, 662

\bibitem[{{Sofia} {et~al.}(1997){Sofia}, {Cardelli}, {Guerin}, \&
  {Meyer}}]{Sofi97}
{Sofia}, U.~J., {Cardelli}, J.~A., {Guerin}, K.~P., \& {Meyer}, D.~M. 1997,
  \apjl, 482, L105+

\bibitem[{{Soifer} {et~al.}(1984){Soifer}, {Neugebauer}, {Helou}, {Lonsdale},
  {Hacking}, {Rice}, {Houck}, {Low}, \& {Rowan-Robinson}}]{Soif84}
{Soifer}, B.~T., {Neugebauer}, G., {Helou}, G., {Lonsdale}, C.~J., {Hacking},
  P., {Rice}, W., {Houck}, J.~R., {Low}, F.~J., \& {Rowan-Robinson}, M. 1984,
  \apjl, 283, L1

\bibitem[{{Solomon} {et~al.}(1992){Solomon}, {Downes}, \& {Radford}}]{Solo92}
{Solomon}, P.~M., {Downes}, D., \& {Radford}, S.~J.~E. 1992, \apjl, 387, L55

\bibitem[{{Solomon} {et~al.}(2004){Solomon}, {Vanden Bout}, \&
  {Maddalena}}]{Solo04}
{Solomon}, P.~M., {Vanden Bout}, P.~A., \& {Maddalena}, R. 2004, Bulletin of
  the American Astronomical Society, 36, 824

\bibitem[{{Sosa-Brito} {et~al.}(2001){Sosa-Brito}, {Tacconi-Garman}, {Lehnert},
  \& {Gallimore}}]{Sosa01}
{Sosa-Brito}, R.~M., {Tacconi-Garman}, L.~E., {Lehnert}, M.~D., \& {Gallimore},
  J.~F. 2001, \apjs, 136, 61

\bibitem[{{Spaans} \& {Meijerink}(2005)}]{Spaa05}
{Spaans}, M. \& {Meijerink}, R. 2005, \apss, 295, 239

\bibitem[{{Umeda} \& {Nomoto}(2002)}]{Umed02}
{Umeda}, H. \& {Nomoto}, K. 2002, \apj, 565, 385

\bibitem[{{van der Tak} {et~al.}(2008){van der Tak}, {Aalto}, \&
  {Meijerink}}]{VanDerTak08}
{van der Tak}, F.~F.~S., {Aalto}, S., \& {Meijerink}, R. 2008, \aap, 477, L5

\bibitem[{{Wagg} {et~al.}(2005){Wagg}, {Wilner}, {Neri}, {Downes}, \&
  {Wiklind}}]{Wagg05}
{Wagg}, J., {Wilner}, D.~J., {Neri}, R., {Downes}, D., \& {Wiklind}, T. 2005,
  \apjl, 634, L13

\bibitem[{{Wei{\ss}} {et~al.}(2007){Wei{\ss}}, {Downes}, {Neri}, {Walter},
  {Henkel}, {Wilner}, {Wagg}, \& {Wiklind}}]{Weis07}
{Wei{\ss}}, A., {Downes}, D., {Neri}, R., {Walter}, F., {Henkel}, C., {Wilner},
  D.~J., {Wagg}, J., \& {Wiklind}, T. 2007, \aap, 467, 955

\bibitem[{{Wu} {et~al.}(2005){Wu}, {Evans}, {Gao}, {Solomon}, {Shirley}, \&
  {Vanden Bout}}]{Wu05}
{Wu}, J., {Evans}, II, N.~J., {Gao}, Y., {Solomon}, P.~M., {Shirley}, Y.~L., \&
  {Vanden Bout}, P.~A. 2005, \apjl, 635, L173

\bibitem[{{Younger} {et~al.}(2008{\natexlab{a}}){Younger}, {Dunlop}, {Peck},
  {Ivison}, {Biggs}, {Chapin}, {Clements}, {Dye}, {Greve}, \&
  {Hughes}}]{Youn08b}
{Younger}, J.~D., {Dunlop}, J.~S., {Peck}, A.~B., {Ivison}, R.~J., {Biggs},
  A.~D., {Chapin}, E.~L., {Clements}, D.~L., {Dye}, S., {Greve}, T.~R., \&
  {Hughes}, D.~H. 2008{\natexlab{a}}, \mnras, 387, 707

\bibitem[{{Younger} {et~al.}(2008{\natexlab{b}}){Younger}, {Fazio}, {Wilner},
  {Ashby}, {Blundell}, {Gurwell}, {Huang}, {Iono}, {Peck}, {Petitpas}, {Scott},
  {Wilson}, \& {Yun}}]{Youn08a}
{Younger}, J.~D., {Fazio}, G.~G., {Wilner}, D.~J., {Ashby}, M.~L.~N.,
  {Blundell}, R., {Gurwell}, M.~A., {Huang}, J.-S., {Iono}, D., {Peck}, A.~B.,
  {Petitpas}, G.~R., {Scott}, K.~S., {Wilson}, G.~W., \& {Yun}, M.~S.
  2008{\natexlab{b}}, ArXiv e-prints

\bibitem[{{Zaritsky} {et~al.}(1994){Zaritsky}, {Kennicutt}, \&
  {Huchra}}]{Zari94}
{Zaritsky}, D., {Kennicutt}, R.~C., \& {Huchra}, J.~P. 1994, \apj, 420, 87

\end{thebibliography}

\end{document}